\documentclass[runningheads]{llncs}
\usepackage{graphicx}
\usepackage{multirow}
\begin{document}
\title{New Hybrid Techniques for Business Recommender Systems}
\author{Charuta Pande \orcidID{0000-0001-6530-5401} \and Hans Friedrich Witschel \orcidID{0000-0002-8608-9039} \and
Andreas Martin \orcidID{0000-0002-7909-7663}}
\authorrunning{Pande et al.}
\institute{FHNW University of Applied Sciences and Arts Northwestern Switzerland, Intelligent Information Systems Research Group, Riggenbachstrasse 16, 4600, Olten, Switzerland\\
\email{\{charuta.pande,hansfriedrich.witschel,andreas.martin\}@fhnw.ch}}
\maketitle              % typeset the header of the contribution
\begin{abstract}
Besides the typical applications of recommender systems in B2C scenarios such as movie or shopping platforms, there is a rising interest in transforming the human-driven advice provided e.g. in consultancy via the use of recommender systems. We explore the special characteristics of such knowledge-based B2B services and propose a process that allows to incorporate recommender systems into them. We suggest and compare several recommender techniques that allow to incorporate the necessary contextual knowledge (e.g. company demographics). These techniques are evaluated in isolation on a test set of business intelligence consultancy cases. We then identify the respective strengths of the different techniques and propose a new hybridisation strategy to combine these strengths. Our results show that the hybridisation leads to a substantial performance improvement over the individual methods.\let\thefootnote\relax\footnote{This article is an extended version of the peer-reviewed publication by Witschel and Martin \cite{Witschel2018RandomWO} and comprises parts from the MSc thesis of the first author Pande \cite{mastersthesis}.}

\end{abstract}
\section{Introduction}
Digitalisation leads to a transformation of internal business processes, but also very notably of customer-facing services. While most attention is paid to services in the B2C domain, there is also a rising interest in digitalising knowledge-intensive services in the B2B domain, such as consultancy in general \cite{nissen2018digital} and IT consultancy in particular \cite{werth2016self}. Such transformation implies that a digital service takes over (partially) the role of a human consultant and that companies can use that service to help themselves to the required advice.

Obviously, such digital services will be able to give advice only for restricted domains -- often, advice will consist in recommending items from a predefined set of solution components. Thus, digital consulting services can be thought of as recommender systems.

As we have laid out in our previous work \cite{Witschel2018RandomWO}, a recommender that suggests solution components to companies are different in several respects from the typical B2C recommenders that help users in finding e.g. books, movies or music that fits their preferences (see also \cite{felfernig2008constraint}):
\begin{itemize}
	\item \textbf{Requirement-driven:} A consultancy recommender needs to consider business requirements, not personal preferences
	\item \textbf{Interdependent items:} The recommended items are not simple, atomic and independent products (such as books, movies etc.), but interdependent and sometimes complex components of a larger solution.
	\item \textbf{No profiles:} While typical B2C recommenders are used repeatedly by the same person, a digital consultancy service has no chance to build up customer profiles through repeated interactions -- companies will usually access the service only once. Thus, a profile of the company needs to be acquired within a single session by the recommender -- one can regard it as forming a \emph{query} that describes the situation of the company seeking advice.
\end{itemize}

\begin{figure}[!h]
	\begin{center}
		\includegraphics[width=0.9\textwidth]{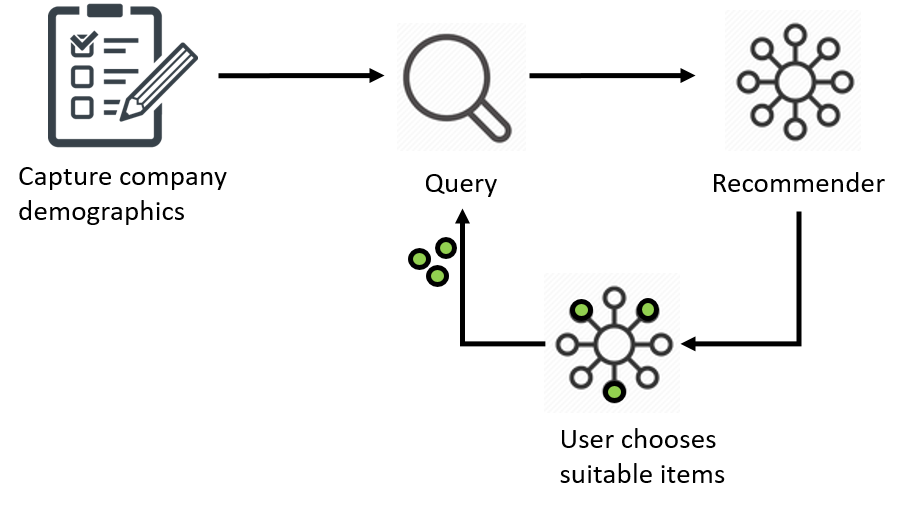}
		\caption{Iterative process for business consultancy recommenders}
		\label{fig:process}
	\end{center}
\end{figure}

Despite some of these differences, one can establish a ``digital consultancy process'' that will make it possible to apply traditional recommender techniques that have been designed for classical preference-based B2C scenarios. Such a process is based on the following considerations (see also Figure \ref{fig:process}):
\begin{enumerate}
	\item Many companies share the same requirements, just like many persons share preferences. The similarity of requirements often depends on the companies' demographics (e.g. size, industry etc.). Thus, a first step in the digital consultancy process may be to capture company demographics and regard them as an initial company profile or \textbf{initial query}. This allows from the beginning to establish a certain similarity between companies.
	\item Later, the similarity of context and requirements manifests itself in accepting similar suggestions from the recommender. Since solutions will be complex, one may construct a repeated interaction with the recommender in the form of iterations: after entering the company demographics (step 1), the business user receives a first set of recommendations and selects from those some first elements of a solution. These elements are added to the initial company profile to form an \textbf{extended query} and the recommender is invoked again. This process is repeated, each time with a more verbose query (we will later use the term ``query verbosity'' to refer to the growing amount of information that the query contains).
\end{enumerate}

Following this iterative process will allow us to assess the \emph{similarity} of company contexts by comparing queries of a company to those of previous users of the service -- with an increasing degree of accuracy as the query is iteratively extended. Since similarity is at the heart of both content-based and collaborative filtering approaches \cite{bobadilla2013recommender}, being able to assess similarities is an important prerequisite for applying these approaches. In addition, we build up company profiles during the process which makes it possible to apply content-based filtering. The iterative refinement makes it also possible to take into account the interdependence of solution elements by identifying, in each new step, new elements that fit to the already selected elements.

Although collaborative and content-based filtering become applicable through our iterative process, they may not be the best choice because a) collaborative filtering does not lend itself readily to incorporating company demographics (or other forms of general context) and b) because both do not foresee the use of human-provided knowledge about the business domain which might be helpful. In fact, previous research has argued for the use of case-based reasoning (CBR) in business recommenders \cite{bridge2005case} because CBR is a proven way of re-using solutions to business problems. Constraint-based recommenders \cite{felfernig2008constraint} are another family of algorithms that have been put forward as a good way of satisfying business requirements. 

In our previous work \cite{Witschel2018RandomWO}, we used a graph, as a simple, flexible and easily extensible means of representing both historic user choices and explicit human knowledge about a business domain, together with a random walk approach to generate recommendations. We found that especially explicit knowledge about associations between solution elements improves the recommender performance. On the other hand, taxonomic knowledge, e.g. about relationships between industries, did not help.

While our previous work was able to benefit from a graph's flexibility and ease of incorporating new domain knowledge easily \cite{Witschel2018RandomWO,minkov2017graph}, it is not perfectly suited to accommodate and make use of all possibly relevant attributes of a company's context. For instance, it does not easily allow to represent and compare numeric attributes (such as company size) or simple string attributes containing longer passages of text. Thus, the goal of our extended research was to explore options of combining graph-based random walks with other forms of recommenders, above all CBR-based ones.

\subsection{Application scenario}
\label{sec:scenario}
We performed our recommender experiments in the domain of IT consultancy, more precisely business intelligence (BI) consultancy. Typically, companies using a BI consultancy service -- before being able to tackle the ``technical'' elements of a BI solution -- initially seek advice regarding

\begin{itemize}
    \item Suitable \emph{key performance indicators (KPIs)} that can be used to monitor and measure the company's success in achieving its goals. A typical KPI might be ``sales revenue''.
    \item Adequate \emph{dimensions} to describe the values of KPIs, e.g. to characterise sales by product that was sold, channel through which it was sold and/or date when it was sold
    \item Suitable \emph{representations}, e.g. charts or tables that help to analyse KPI values along dimensions (e.g. a chart showing temporal evolution of sales revenue for different products)
\end{itemize}
Here, we focus on the first two types of solution elements. Obviously, the question of which dimensions should be chosen depends on the KPIs to be monitored. Choice of KPIs, in turn, is often determined by the (type of) industry of a company -- e.g. companies who produce energy tend to have KPIs that differ substantially from those of, say, architects. KPIs are also usually determined by the business process (e.g. sales) that should be analysed. 
%Finally, companies are usually able to express specific goals for their future analytical solution.

\subsection{Contribution}
\label{sec:contribution}
Given the application scenario that we just sketched, our main research goal in this work is to construct a new hybrid recommender that optimally supports the requirements of a B2B consultancy service. We investigate hybridisation because
\begin{enumerate}
    \item despite the existence of some previous work, we do not yet have reliable knowledge about which type of recommender is best suited for the task and
    \item we do know that different recommenders have different strengths and weaknesses in general that affect their ability to represent and accommodate certain types of knowledge and/or inputs and their ability to deal with lack of such knowledge (``cold-start problems'').
\end{enumerate} 

We will first investigate the performance of algorithms individually and then -- by a more detailed analysis of their strengths and weaknesses on the data -- propose and evaluate some hybridisation strategies that will lead to superior performance by joining the strengths of the best-suited recommenders.

\section{Related Work}
\label{sec:related}
Digitalising consultancy services has been discussed recently for the domain of IT consulting. In \cite{werth2016self}, a ``computer-executed consulting (CEC) service'' is proposed, which replaces, most notably, the two steps of a) interviewing client representatives and b) creating a report that summarises the interview results. The digital service is designed by human consultants and consists of a) a series of questionnaires (replacing the interviews) and b) an automated report creation module. Obviously, there is a rough correspondence between these components and the step of a) formulating a query and b) getting recommendations for that query in Figure \ref{fig:process}. The proposed CEC service is general-purpose. Therefore, although it mentions the need for more intelligence in the report creation module and the option of using recommender systems, it does not discuss any details of how to use recommenders.\\

Application of recommender systems has been discussed for more specific consultancy tasks such as optimisation of product assortments \cite{witschel15c}, selection of cloud or web services \cite{zhang2012declarative,kritikos2017towards,yao2015unified} or adaptation of conditions in agriculture \cite{laliwala2006semantic}. In all these cases, the set of possible items that can be recommended is known and well-defined and the task consists in selecting and possibly orchestrating the items. In its simplest interpretation, the term ``orchestration'' means simply that the selected services should be well aligned with each other, e.g. for optimal cross-selling opportunities \cite{witschel15c} or for obtaining a consistent complex cloud service configuration \cite{yao2015unified}. This is also the case for our BI consultancy service, see Section \ref{sec:scenario} and \cite{Witschel2018RandomWO}.\\

In terms of algorithms, business-oriented recommender systems have to deal with \textbf{complexity} in terms of company contexts (input) and solutions (output). Attempts to deal with such complexity can be divided into several categories:
\begin{itemize}
    \item Augmentations of content-based filtering (CBF): approaches in this category model both the input and the output complexities and establish the degree to which both of them match. For instance, constraint-based recommenders \cite{felfernig2008constraint,felfernig2015constraint} help to model product features and constraints to be expressed about them and then ensure constraint satisfaction. Other approaches use tree-like structures to model items and user preferences \cite{wu2015fuzzy} or use multiple levels on which queries and items are matched (such as recommending first providers and then actual services in a service recommender, \cite{mohamed2016multi}).
    In CBF, additional knowledge can be incorporated, e.g. into the function that determines the similarity between an item and the user profile. Often, this is knowledge about user context, item features and/or domain-specific constraints. For instance, \cite{carrer2012social} and \cite{blanco2008flexible} use ontologies to represent and reason about item features and to apply this knowledge in a sophisticated similarity measure that takes into account ``hidden relationships'' \cite{blanco2008flexible}. Middleton et al. \cite{middleton2004ontological} use an ontology to represent user profiles and engage users in correcting the profiles before assessing profile-item similarities.
    \item Augmentations of collaborative filtering: Case-based recommenders \cite{bridge2005case,bousbahi2015mooc} can be seen as a special form of collaborative filtering since they recommend items used in solutions of companies that are similar to the current company. However, instead of only considering already chosen items, case-based recommenders' similarity measures take into account context variables that describe e.g. company demographics and other relevant aspects of the company's problem and/or initial situation.
    \item Graph-based recommenders \cite{bogers2010movie,zhang2013random,minkov2017graph} have been put forward because of their ability to accommodate a wide variety of forms of contexts in a flexible way without much effort. Random walks \cite{Fouss2007,Huang2002} are a predominant type of algorithm to provide recommendations based on graph structures. Because of their simplicity, graphs also have limitations, e.g. in modeling and matching simple string-valued attributes of input cases or in modeling certain forms of complex solution structures. The possibility to use graph-based recommenders to ``mimick'' traditional recommender approaches, such as collaborative or content-based filtering, has been explored in \cite{lee2013pathrank}. For this, one needs to assign different weight to different types of graph relations. 
\end{itemize}

Obviously, all of these approaches employ and model various types of knowledge. An overview of the different kinds of knowledge that recommenders may use can be found in \cite{felfernig2008constraint,Witschel2018RandomWO}. What distinguishes the business recommenders from most others is the use of \emph{domain knowledge}. Often, this knowledge is obtained from human experts as discussed in \cite{felfernig2008constraint,tarus2017knowledge,Witschel2018RandomWO}.\\

Finally, forming hybrid recommenders \cite{burke2002hybrid,burke2007hybrid} is an active field of research since combinations of different approaches can often help to combine the strengths and/or avoid the weaknesses of the combined approaches. For instance, content-based filtering can be combined with collaborative filtering (CF), e.g. to mitigate the so-called cold-start problems associated with CF, i.e. problems with recommending newly introduced items or serving new users: new items can be recommended immediately by content-based techniques as long as they have a meaningful description that can be matched against user profiles. 
Besides cold start problems, hybridisation can be used e.g. to augment similarity in collaborative filtering with the reasons behind user preferences and thus give it a stronger CBR flavour \cite{burke2000case}. Further possibly complementary strengths and weaknesses of knowledge-based and knowledge-weak recommenders are discussed in \cite{burke2000knowledge}.\\

Overall, there is a rather large number of suggestions for enriching recommenders with contextual knowledge. However, as outlined in Section \ref{sec:contribution}, we see a gap in exploring which of these suggestions is best suited to support scenarios of business consultancy. We furthermore see a need to gain a deeper understanding of the (complementary) strengths and weaknesses of the mentioned approaches that will lead to successful hybridisation strategies.

\section{Methodology}
%\begin{itemize}
%    \item interviews with consultants, building the case base: see old KMIS paper
%    \item leave-one-out experiments and pooling experiment with new case: see Charuta's MT
% \end{itemize}
 
As mentioned in Section \ref{sec:contribution}, the main goal of our research is to find a recommender form that optimally supports B2B consultancy services. To study such services, we worked together with a company that provides business intelligence (BI) consultancy, as described in Section \ref{sec:scenario}.

\subsection{Awareness of current consultancy practice}
\label{sec:method-awareness}
As described in our previous work \cite{Witschel2018RandomWO}, our research started by interviewing two consultants to understand how they work and which knowledge they require to make the necessary recommendations to their customers. We also obtained some documents that were used to document the outcomes of meetings and workshops with customers. 

This was the basis for us to define the structure of consultancy cases: it gave us an insight into the demographic and contextual variables (attributes) that consultants need to know about each company. It also allowed us to grasp roughly the kind of reasoning that they employed to transfer their experiences to new cases. The corresponding findings are summarised in Section \ref{sec:interviews}.\\

%TODO: move the result part of this to the new Section 4
We then constructed a case base out of the past experience of the consultancy and identified cases that represent the business context of customers; each business process that a company wanted to analyse resulted in a separate case. Overall, this resulted in a case base with 82 entries.\\

To support our extended research, we performed a second round of interviews to gain further awareness of how consultants currently assess (implicitly or explicitly) the similarity between customer cases. More precisely, we asked them to which degree they take into account each attribute in the case (e.g. the industry, the core business processes to be analysed, the target group, the goal of the consultation), i.e. we elicited the importance they assign to each attribute while deriving recommendations for their customers.

\subsection{Recommender selection and configuration}
Next, we used the gathered knowledge to configure a selection of recommender algorithms that we wanted to compare:
\begin{itemize}
	\item Collaborative filtering, using both item-based and user-based k-nearest neighbour algorithms, as provided by the LibRec library \cite{guo2015librec}.
	\item A random walk algorithm based on a ``case graph'' as described in \cite{Witschel2018RandomWO}.
	\item A CBR-based recommender that applies similarity-weighted scoring to the elements contained in similar cases. The weights mentioned above were used here to define the contribution of the 	local similarities within the global similarity function in CBR.
\end{itemize}
A precise description of recommender configurations can be found in Section \ref{sec:config}.

\subsection{Experiments}
We then designed an experimental setup \cite{mastersthesis} to compare the initial recommender configurations, as well as our new hybrid recommender strategies. 

This setup consists in a \textit{leave-one-case-out} evaluation: for each case $C$, we used the case base as the training data by \textit{omitting} $C$. Out of $C$, we constructed queries $Q_C$ at different verbosity levels: simple queries with no input elements and gradually more verbose queries containing an increasing number of randomly chosen KPIs from the case $C$. 

The random selection of the input elements is not realistic as this information is usually provided by the customer. However, we did not have any information about the order in which customers added elements to their solution in the past and thus had to resort to this strategy.

For the evaluation of recommender outputs, we used the knowledge of originally chosen elements in $C$ as a definition of relevance: for a query $Q_C$, we observed whether a recommender was able to retrieve (and rank highly) the elements in the original case $C$. That is, for each \emph{ranking} of recommended items that a recommender produced in response to a query $Q_C$, we computed mean average precision \cite{voorheesBook} of these rankings by treating all elements originally contained in $C$ as relevant and all others as irrelevant.

As mentioned above, this experimental setup was used first to evaluate each recommender in isolation. We then analysed the strengths and weaknesses of each recommender (see Section \ref{sec:strengths}) and formed new hybrid recommender strategies (see Section \ref{sec:hybrid}) that we evaluated with the same experimental environment to see whether the hybridisation could bring about an improvement (see Section \ref{sec:experiment2}).

\section{Interview findings: case structure and similarity measure}
\label{sec:interviews}
As mentioned in Section \ref{sec:method-awareness}, we performed two rounds of interviews with consultants to understand their current work and knowledge processing procedures. Here, we summarise the findings from both rounds of interviews (see also \cite{Witschel2018RandomWO} for more details on the first round):
\begin{itemize}
	\item Customers often come to the meetings with some important KPIs and dimensions (i.e. solution elements) already in mind. However, the degree to which customers have initial ideas can vary greatly. We have reflected this variance by creating queries at different verbosity levels.
	\item In terms of company demographics, consultants consider the industry of a customer as the main criterion for finding similar past cases. Further relevant variables that we elicited were the target group of the solution (e.g. only management or all employees) and the goal of the BI project (expressed in natural language). Finally, consultants use of course all known customer preferences from initial meetings (see above), i.e. any already known solution elements to remember past cases with similar elements. 
	
	The business process was also mentioned by consultants as an important variable. Because of its importance, we chose not to use it simply as a ranking criterion for the retrieval of similar cases, but as a filter: for a given company, we created separate cases for each business process the company wanted to analyse and retrieved only cases with the same business process (analogously, we built separate case base graphs for the graph recommender, see below).
	% consultants also consider core/support process as the main criteria besides industry. I think we should mention it, especially since I filtered out cases that do not match the process of the query case in CBR and there was a significant improvement in MAP
	\item In the second round of interviews, we asked the consultants to quantify the relative importance of these types of attributes. Although quantifying something as abstract as a variables contribution to a similarity score is a hard task, we were able to verify in some preliminary experiments that the chosen weights gave quite good results as compared to other potential weight configurations. The resulting weights are shown in Table \ref{tab:weights}.
	\item When talking to a customer from a yet unknown industry, consultants tried to remember cases of customers from \emph{similar} industries. Since our attempts to use an industry taxonomy for improved similarity assessment in a graph-based recommender were not particularly successful, we did not consider this kind of reasoning in this work. However, we did use the industry taxonomy to define a local similarity measure for industries within a CBR-based recommender (see Section \ref{sec:cbr-rec}).
\end{itemize}

\begin{table}
	\caption{Local similarities and weights for CBR recommender}
	\begin{center}
		\begin{tabular}{ p{3cm} | p{3cm} | p{2cm} }
			\textbf{Case Attribute} & \textbf{Local similarity measure} & \textbf{Weight} \\
			\hline
			Industry & Taxonomy & 0.24 \\ 
			\hline
			Goal & TF-IDF & 0.06 \\
			\hline
			Target Group & Jaccard & 0.1 \\
			\hline
			KPIs and dimensions & TF-IDF & 0.6 \\
			
		\end{tabular}
	\end{center}
	\label{tab:weights}
\end{table}

\section{Recommender configurations}
\label{sec:config}
Based on the interview findings, we created suitable configurations of the  recommenders to be used in the experiments \cite{mastersthesis}, as described in the following subsections.

\subsection{Collaborative filtering}
Since the association between solution elements (which we will call items for simplicity) and cases is binary -- an item is either part of the case's solution or not -- we can describe this situation as one of ``implicit feedback recommendation'' \cite{zhang2018social}. It means that the user-item matrix does not contain true ratings, but binary entries -- in our case, we replaced users with customers. 

However, this does not require to change the way in which Collaborative Filtering algorithms work on the matrix. In our experiment, we used the user-based \emph{userknn} and the item-based \emph{itemknn} implementations from the LibRec package \cite{guo2015librec}.\\

Since \emph{userknn} and \emph{itemknn} do not allow us to make use of the additional attributes listed in Table \ref{tab:weights}, ``simple'' queries that do not contain any items (verbosity level 0) %do not return any results% 
could not be designed. We also expect the collaborative filtering algorithms to have inferior results for low verbosity queries.

\subsection{Random walk recommender}    
The configuration for the graph-based recommender was re-used from \cite{Witschel2018RandomWO}, where the case graph incorporated the explicit knowledge acquired from the consultants. In this technique, the case graph was built by creating a node for each case and connecting it to a node representing the industry as well as to nodes representing solution elements. As mentioned in Section \ref{sec:interviews}, we built a separate graph for each business process to be analysed.

Target group and goal were not represented in this approach: since there are only three possible target groups, the corresponding nodes would have had a very high degree, thus diluting the PageRank scores. Since goals are string attributes, a node representation was not straightforward for them (although future work might consider extracting salient terms and representing them as nodes).

The recommended elements were scored using the PageRank with Priors algorithm \cite{White2003} on that graph. The scores represent the probability of reaching a node in the case graph (e.g. the elements to be recommended) through a random walk that is biased towards the input elements in the query.

For verbosity level 0, the random walk-based recommender uses only the industry node as a query -- we also expect suboptimal results here.

\subsection{CBR-based recommender}
\label{sec:cbr-rec}
In the case of the CBR recommender, primarily three factors were considered in the configuration: 
\begin{itemize}
\item Similarity measures depending on attribute type: based on the taxonomy-tree approach proposed by \cite{bergmann1998use}, the industry attribute uses the industry taxonomy derived by \cite{Witschel2018RandomWO} that categorizes the customers of the consultancy based on their similarities (e.g. customers that are likely to share KPIs and dimensions). For the attributes goal (free text) and KPI, we could apply the TF-IDF \cite{huang2008similarity} similarity measure by creating a corpus of goals and KPIs respectively from the case base for the computation of inverse document frequencies (IDF). Although KPIs are not free text, applying TF-IDF is appropriate to disregard repeated terms like "Number of", "Amount", since they do not add significant value to the recommendations. Lastly, for the attribute target group, we calculated the Jaccard coefficient \cite{huang2008similarity} as a case may have more than one target audience from the possible values "employees"/"middle management"/"top management".
\item The number $n$ of the most relevant (top) cases retrieved: The number of the retrieved cases played a significant role in calculating the scores of the recommended elements, which in turn determine the ranking. For an element appearing in any of the retrieved cases $R(Q_C)$ for a query $Q_C$, the score of that element is the sum of the scores of all the retrieved cases in which the element occurs:
\begin{equation}
score(i) = \sum_{C_j \in R(Q_C):i \in C_j}{sim(C,C_j)}
\end{equation}
Obviously, the larger the case base, the larger we can choose $n$, i.e. the maximum size of $R(Q_C)$. For a rather small case base like ours, we expect that smaller values of $n$ will work better since larger values will likely imply a ``topic drift'' by including rather dissimilar cases.
The score of the case $sim(C,C_j)$ was generated by the CBR recommender using the global similarity function, which is the weighted average of the local similarity measures \cite{Richter1998}: $sim(C,C_j) = \sum{w_k sim_k(C,C_j)}$. 
\item For that weighted average, we used the weights assigned to the local similarity measures $sim_k$ shown in Table \ref{tab:weights}.
\end{itemize}
The retrieved ranking of matching cases was %finally 
first filtered by business process such as to return only cases with matching process before applying the local similarity measures.
% mention that only cases that matched the process of the query case were considered for recommendation -> done

\section{Experiment 1: Strengths and weaknesses of recommenders}
\label{sec:strengths}
%\begin{itemize}
%    \item Summary of Charuta's MT experiments
%    \item Conclusion: a hybrid strategy is needed
%\end{itemize}
The goal of our first experiment was to identify a recommendation technique that performs well for different query verbosity values \cite{mastersthesis}. The results of Experiment 1 are shown in Table 2. Note that the verbosity refers to the absolute number of solution elements that the query contained. 

From the results, we can see very clearly that the Collaborative Filtering algorithms obviously suffer too much from their inability to accommodate contextual knowledge. Their performance is substantially worse than that of the other recommenders.

%We started with the CBR configuration to retrieve a single case most similar to the query. 
Regarding, those, we observed that the performance of the CBR recommender is better than the graph-based recommender, however, there is no improvement in the performance of the CBR recommender above a certain query verbosity. Thus, one can see that retrieving a single case is restrictive for the recommendations since only a limited number of elements are available which in turn creates a recall problem. The performance of the graph-based recommender, on the other hand, steadily improves as more elements are added to the query. In order to enable the CBR recommender to stretch its (better) performance to any size of the query, we repeated the leave-one-case-out evaluation by retrieving more number of most relevant cases. With the top two retrieved cases, the performance of the CBR recommender improved further, however, again only up to a certain query verbosity. By retrieving more and more cases, it was possible to overcome the recall problem and achieve a steady improvement in the performance of the CBR recommender, similar to the graph-based recommender. Nonetheless, one can observe that retrieving more cases also introduces more noise, consequently decreasing the overall performance of the CBR recommender. Thus, increasing the number of retrieved cases seems to be neither the optimal nor a generic solution to the recall problem of the CBR recommender because of its severe precision-degrading effect.

\begin{table}
\caption{Experiment 1: MAP values for individual recommendation techniques for different configurations}
\begin{center}
\begin{tabular}{ p{1.6cm} | p{1.1cm} | p{1.1cm} | p{1.5cm} | p{1.5cm} | p{1.5cm} | p{1.5cm}| p{1.5cm}}
\multirow{2}{10em}{\textbf{Query\\verbosity}} & \textbf{user-knn} & \textbf{item-knn} & \textbf{Graph-based} & \multicolumn{4}{c}{\textbf{CBR}}\\
{} & {} & {} & {} & {top 1} & {top 2} & {top 3} & {top 5} \\
\hline
 0 & - & - & 0.408 & 0.773 & 0.783 & 0.773 & 0.747 \\ 
 5 & 0.487 & 0.420 & 0.566 & 0.777 & 0.805 & 0.774 & 0.714 \\ 
 10 & 0.497 & 0.416 & 0.646 & 0.785 & 0.805 & 0.772 & 0.719 \\ 
 15 & 0.498 & 0.413 & 0.689 & 0.787 & 0.807 & 0.766 & 0.709 \\ 
 20 & 0.501 & 0.411 & 0.713 & 0.787 & 0.807 & 0.776 & 0.713 \\ 
 30 & 0.498 & 0.411 & 0.733 & 0.787 & 0.812 & 0.780 & 0.716 \\ 
 40 & 0.499 & 0.411 & 0.742 & 0.787 & 0.812 & 0.783 & 0.719 \\ 
 100 & 0.499 & 0.409 & 0.746 & 0.787 & 0.812 & 0.781 & 0.718 \\ 
% 0 & ?? & ?? & 0.408 & 0.773 & 0.783 & 0.773 & 0.754 & 0.747 \\ 
% 5 & ?? & ?? & 0.566 & 0.777 & 0.805 & 0.774 & 0.730 & 0.714 \\ 
% 10 & ?? & ?? & 0.646 & 0.785 & 0.805 & 0.772 & 0.735 & 0.719 \\ 
% 15 & ?? & ?? & 0.689 & 0.787 & 0.807 & 0.766 & 0.722 & 0.709 \\ 
% 20 & ?? & ?? & 0.713 & 0.787 & 0.807 & 0.776 & 0.725 & 0.713 \\ 
% 30 & ?? & ?? & 0.733 & 0.787 & 0.812 & 0.780 & 0.729 & 0.716 \\ 
% 40 & ?? & ?? & 0.742 & 0.787 & 0.812 & 0.783 & 0.731 & 0.719 \\ 
% 100 ?? & ?? & & 0.746 & 0.787 & 0.812 & 0.781 & 0.731 & 0.718 \\ 
\end{tabular}
\end{center}
\end{table}

Overall, to achieve an optimum performance, the CBR recommender needs to be configured to retrieve a low number of most relevant cases. Yet, if a customer needs a solution with more elements than are available in the (small number of) retrieved cases, the CBR recommender fails to expand its range of recommendations. The graph-based recommender, on the other hand, can leverage the whole range of elements available in the case base and hence seems to be a better solution for increasing recall without adding too much noise. We, therefore, see a benefit in combining the graph-based and CBR recommendation techniques using a hybrid strategy.

\section{A new hybrid recommender design}
\label{sec:hybrid}
%Describe the two alternative forms of hybridisation that we explored, their (a priori) pros and cons - just the one considering verbosity

\begin{figure}[ht]
\centering
\includegraphics[scale=0.3]{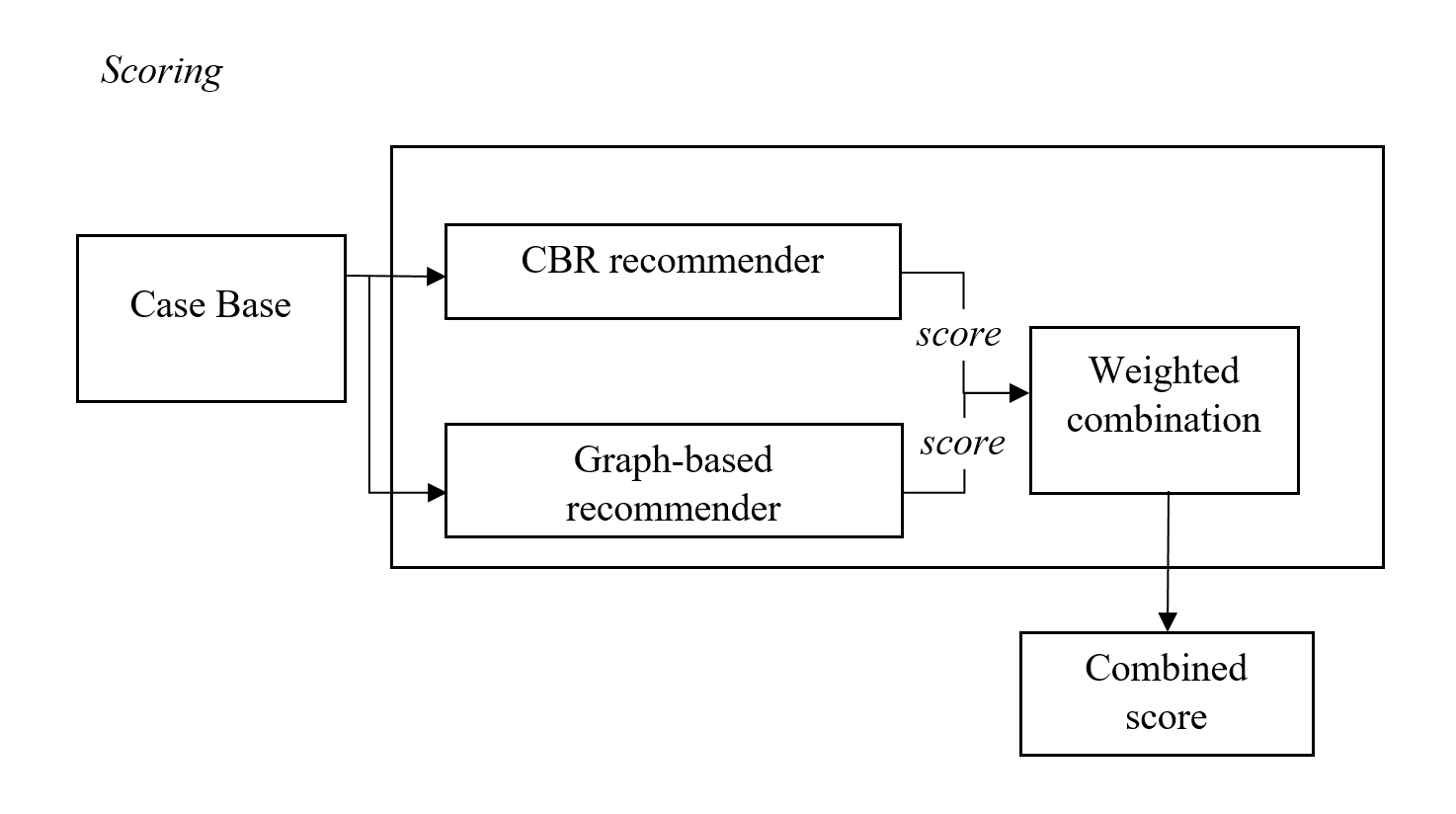}
\caption{Weighted Hybrid design, adapted from \cite{burke2007hybrid}}
\label{fig:Hybrid}
\end{figure}

In Section \ref{sec:related}, we saw that Hybrid recommender systems are commonly used to overcome the weaknesses of individual recommendation techniques. Of the seven hybrid recommender strategies described by \cite{burke2007hybrid},  strategies like \textit{switching}, \textit{cascade} or \textit{mixed} are not ideal (and the others are not applicable), as the results show that CBR recommender is clearly the better performer.  Since we would like the graph-based recommender to contribute by adding more relevant elements where CBR is limited, we adopted the weighted combination method because it allows to "overrule" the decisions of the CBR by adjusting the importance (weight) given to either CBR or graph-based recommender. A representation of the weighted hybrid strategy adopted by us is shown in Figure \ref{fig:Hybrid}.

For designing the hybrid strategy, we built upon the CBR configuration to retrieve the most relevant two cases, as this configuration achieved the optimum performance in the previous experiment. We now explore if the recall issue of CBR can be resolved by adding some component of the graph-based recommendations. We first normalised the scores of the individual recommendation techniques using min-max normalisation, since the graph-based and CBR recommenders have their own (different) scoring mechanisms, as described in Section \ref{sec:config}. We then combined the normalised scores of both recommenders and calculated the hybrid weighted score using Equation \ref{eq:hybridscore}. 

\begin{equation}
\label{eq:hybridscore}
    hybrid\_score(item)=\alpha \cdot |CBR(item)| + (1-\alpha) \cdot |PR(item)|
\end{equation}

where $|\cdot|$ refers to min-max score normalisation.\\

Because of the CBR recommender's strength in dealing with sparse, i.e. low-verbosity query and the relative strength of the graph-based recommender in handling high-verbosity queries, we made the mixture parameter $\alpha$ dependent on the query verbosity, i.e. the number of referred elements $|q|$ in the query $q$:
\begin{equation}
\label{eq:alpha}
    \alpha = \left\{ \begin{array}{ll} 1-\frac{(1-\beta)\cdot |q|}{\bar c} & \textrm{if $|q| \leq \bar c$}\\\beta & \textrm{otherwise}\end{array} , 0 < \beta < 1 \right.\\
\end{equation}
Here, $\bar c$ refers to half the average size of all cases in the case base in terms of their number of referred elements (KPIs) and serves as the "verbosity threshold". 

\begin{figure}[ht]
	\centering
	\includegraphics[width=0.6\textwidth]{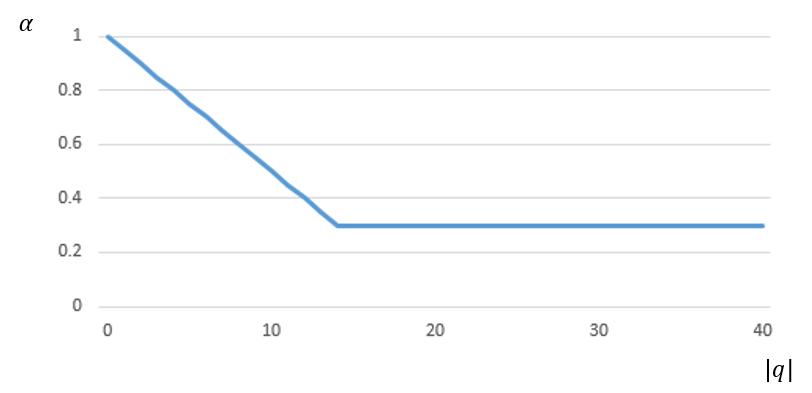}
	\caption{Mixture parameter $\alpha$ as a function of query verbosity $q$}
	\label{fig:alpha}
\end{figure}

Since CBR was the better performer of the two recommendation techniques, we designed Equation \ref{eq:alpha} such that the weight of the CBR recommender ($\alpha$) is never 0. On the other hand, we do not set $\beta$ to 1 as this would give a full weight to CBR, which we already know has limitations performing as a "pure" recommender. Additionally, from the results of Experiment 1, we concluded that a CBR-heavy hybrid recommender would perform better for queries below the verbosity threshold and vice-versa, also taken care in Equation \ref{eq:alpha}. Figure \ref{fig:alpha} shows the dependency between $\alpha$ and query verbosity $|q|$ graphically, for $\beta=0.3$ and $\bar c=14$. We can see how $\beta$ acts as the ``minimum CBR contribution'' and that below the verbosity threshold, less and less weight is given to CBR as verbosity increases.

\begin{table}
\caption{Experiment 2: MAP values for individual recommendation techniques and hybrid strategy}
\begin{center}
\begin{tabular}{ p{2cm} | p{1.5cm} | p{1.5cm} | p{1.5cm} | p{1.5cm}| p{1.5cm} | p{1.5cm}}
\multirow{2}{10em}{\textbf{Query size (verbosity)}} & \textbf{Graph-based} & \textbf{CBR} & \textbf{Hybrid} & \textbf{Hybrid} & \textbf{Hybrid} & \textbf{Hybrid} \\
{} & {} & {top 2 cases} & {$\beta$=0.1} & {$\beta$=0.3} & {$\beta$=0.5} & {$\beta$=0.9} \\
\hline
 0 & 0.408 & 0.783 & 0.801 & 0.801 & 0.801 & 0.801 \\ 
 5 & 0.566 & 0.805 & 0.836 & 0.835 & 0.836 & 0.835 \\ 
 10 & 0.646 & 0.805 & 0.843 & 0.845 & 0.839 & 0.840 \\ 
 15 & 0.689 & 0.807 & 0.804 & 0.856 & 0.849 & 0.849 \\ 
 20 & 0.713 & 0.807 & 0.817 & 0.862 & 0.853 & 0.851 \\ 
 30 & 0.733 & 0.812 & 0.821 & 0.864 & 0.856 & 0.852 \\ 
 40 & 0.742 & 0.812 & 0.825 & 0.866 & 0.858 & 0.854 \\ 
 100 & 0.746 & 0.812 & 0.827 & 0.866 & 0.858 & 0.854 \\ 
 
\end{tabular}
\end{center}
\end{table}

With this setup, we carried out the second experiment - leave-one-case-out evaluation for different query verbosity, using the hybrid strategy. Our goal, now, was to find the appropriate combination of weights that could overcome the recall issue of CBR without impacting its performance. After every run, we compared the mean average precision for each recommendation technique with that of the hybrid strategy, as seen in Table 3.

\section{Experiment 2: Performance of hybrid recommender}
\label{sec:experiment2}
% Hopefully better than each recommender alone...

To find the right combination of the graph-based and CBR recommender, we experimented with different values of $\beta$, starting with a very low value. The lower values of $\beta$ indicate a higher weight to the graph-based recommender. The performance of the hybrid recommender appears to be better than either of the individual recommendation techniques, however, the precision issue of the graph recommender still shows its negative impact for very low values of $\beta$. The verbosity threshold for our experiments was at 14, and it can be observed that the performance suddenly dips at 15 input elements for $\beta$=0.1 (where MAP = 0.843 for a verbosity of 10 and MAP = 0.804 for verbosity 15).\\

On the other hand, although a high $\beta$ resolves the precision problem, the performance is not optimum because for e.g. $\beta=0.9$, the graph recommender's ability to provide more recall is not sufficiently leveraged.

From the results for $\beta$=0.3, one can conclude that the right value of $\beta$ can cure both the recall problem of the CBR recommender and the precision problem of the graph recommender, and thus gives an optimum performance among the individual recommendation techniques and the various configurations of the hybrid strategy put together.

\section{Conclusions}
In this work, we considered the application of recommender systems to business consultancy. We have argued how certain consultancy tasks can be formulated as recommendation problems, especially in the domain of IT consultancy -- e.g. selection and orchestration of web services or selection of Key Performance Indicators and dimensions for Business Intelligence (BI) solutions. Since such problems are in several respects different from the typical, purely preference-based B2C recommenders, we have addressed the question which (combinations of) recommendation techniques are most suitable for these new B2B scenarios.

We worked with data from the BI consultancy domain and performed experiments with a range of known recommender techniques. These techniques offer a varying degree of possibility to feed -- besides the item choices that a company makes -- contextual knowledge, such as company demographics, into the algorithm. This ranges from none (collaborative filtering) over limited (graph-based random walks) to full coverage (CBR-based recommender).

Our initial comparison showed that -- as one might expect -- the CBR-based recommendation benefits from its ability to accommodate more contextual knowledge and provides the best results. However, we also recognised a limitation: CBR-based recommenders have a free parameter, namely $n$, the number of most similar cases to use for the identification of possible solution elements. We found that, for the rather small case base in our experiments, small values of $n$ performed better. Obviously, a larger $n$ implies more noise coming from more dissimilar cases. In our previous work \cite{Witschel2018RandomWO}, we already observed that including cases e.g. from different, but similar industries can be dangerous.

On the other hand, limiting $n$ also limits the potential recall of the recommender, i.e. some useful items from less similar cases are excluded. Obviously, a graph-based approach -- although less precise -- offers a natural way to include more items, also from the more dissimilar cases.\\

We, therefore, explored the combination of CBR-based recommendation with a graph-based recommender in order to combine its strengths in terms of precision with the graph-based recommender's strength in providing more relevant items in the lower ranks. We followed a weighted hybridisation strategy. The weight was dynamic, giving more and more importance to the graph recommender with the growing size of the query. This makes sense since contextual knowledge becomes less important as we know more and more about already chosen items. Because of the superior performance of the CBR recommender, we also designed the weighting so as to ensure that there is always a certain minimum weight given to it. It turned out that indeed this minimum weight should not be 0.\\

We found that the weighted hybrid performed -- at all levels of query verbosity -- better than any of the individual recommenders. Although we have only tested the hybrid on one particular data set, we believe that we can carefully conclude from this that a CBR recommender's problems in balancing between precision and recall can be overcome by combining it with another recommender that is less limited by case boundaries and can contribute better recall at lower ranks. The graph-based recommender was able to achieve that in our experiments.

In future work, we plan to apply our approach also to different domains and data sets. In that context, it will also be important to study more closely the relationship between the size and characteristics of the case base and the optimal choice of the parameter $n$ of the case-based recommender.

% Limitation - we still do not know the strategy to find out the appropriate value of N (we know it should be low)

\bibliographystyle{splncs04}
\bibliography{paper}

\begin{thebibliography}{10}
\providecommand{\url}[1]{\texttt{#1}}
\providecommand{\urlprefix}{URL }
\providecommand{\doi}[1]{https://doi.org/#1}

\bibitem{bergmann1998use}
Bergmann, R.: On the use of taxonomies for representing case features and local
  similarity measures  (1998)

\bibitem{blanco2008flexible}
Blanco-Fern{\'a}ndez, Y., Pazos-Arias, J.J., Gil-Solla, A., Ramos-Cabrer, M.,
  L{\'o}pez-Nores, M., Garc{\'\i}a-Duque, J., Fern{\'a}ndez-Vilas, A.,
  D{\'\i}az-Redondo, R.P., Bermejo-Mu{\~n}oz, J.: A flexible semantic inference
  methodology to reason about user preferences in knowledge-based recommender
  systems. Knowledge-Based Systems  \textbf{21}(4),  305--320 (2008)

\bibitem{bobadilla2013recommender}
Bobadilla, J., Ortega, F., Hernando, A., Guti{\'e}rrez, A.: Recommender systems
  survey. Knowledge-based systems  \textbf{46},  109--132 (2013)

\bibitem{bogers2010movie}
Bogers, T.: Movie recommendation using random walks over the contextual graph.
  In: Proc. of the 2nd Intl. Workshop on Context-Aware Recommender Systems
  (2010)

\bibitem{bousbahi2015mooc}
Bousbahi, F., Chorfi, H.: Mooc-rec: a case based recommender system for moocs.
  Procedia-Social and Behavioral Sciences  \textbf{195},  1813--1822 (2015)

\bibitem{bridge2005case}
Bridge, D., G{\"o}ker, M.H., McGinty, L., Smyth, B.: Case-based recommender
  systems. The Knowledge Engineering Review  \textbf{20}(3),  315--320 (2005)

\bibitem{burke2000case}
Burke, R.: A case-based reasoning approach to collaborative filtering. In:
  European Workshop on Advances in Case-Based Reasoning. pp. 370--379. Springer
  (2000)

\bibitem{burke2000knowledge}
Burke, R.: Knowledge-based recommender systems. Encyclopedia of library and
  information systems  \textbf{69}(Supplement 32),  175--186 (2000)

\bibitem{burke2002hybrid}
Burke, R.: Hybrid recommender systems: Survey and experiments. User modeling
  and user-adapted interaction  \textbf{12}(4),  331--370 (2002)

\bibitem{burke2007hybrid}
Burke, R.: Hybrid web recommender systems. In: The adaptive web, pp. 377--408.
  Springer (2007)

\bibitem{carrer2012social}
Carrer-Neto, W., Hern{\'a}ndez-Alcaraz, M.L., Valencia-Garc{\'\i}a, R.,
  Garc{\'\i}a-S{\'a}nchez, F.: Social knowledge-based recommender system.
  application to the movies domain. Expert Systems with applications
  \textbf{39}(12),  10990--11000 (2012)

\bibitem{felfernig2008constraint}
Felfernig, A., Burke, R.: Constraint-based recommender systems: technologies
  and research issues. In: Proceedings of the 10th international conference on
  Electronic commerce. p.~3. ACM (2008)

\bibitem{felfernig2015constraint}
Felfernig, A., Friedrich, G., Jannach, D., Zanker, M.: Constraint-based
  recommender systems. In: Recommender systems handbook, pp. 161--190. Springer
  (2015)

\bibitem{Fouss2007}
Fouss, F., Pirotte, A., Renders, J.M., Saerens, M.: {Random-Walk Computation of
  Similarities Between Nodes of a Graph with Application to Collaborative
  Recommendation}. IEEE Transactions on Knowledge and Data Engineering
  \textbf{19}(3),  355--369 (2007)

\bibitem{guo2015librec}
Guo, G., Zhang, J., Sun, Z., Yorke-Smith, N.: Librec: A java library for
  recommender systems. In: UMAP Workshops. vol.~4 (2015)

\bibitem{huang2008similarity}
Huang, A.: Similarity measures for text document clustering. In: Proceedings of
  the sixth new zealand computer science research student conference
  (NZCSRSC2008), Christchurch, New Zealand. vol.~4, pp. 9--56 (2008)

\bibitem{Huang2002}
Huang, Z., Chung, W., Ong, T.H., Chen, H.: {A Graph-based Recommender System
  for Digital Library}. In: Proceedings of the 2Nd ACM/IEEE-CS Joint Conference
  on Digital Libraries. pp. 65--73 (2002)

\bibitem{kritikos2017towards}
Kritikos, K., Laurenzi, E., Hinkelmann, K.: Towards business-to-it alignment in
  the cloud. In: European Conference on Service-Oriented and Cloud Computing.
  pp. 35--52. Springer (2017)

\bibitem{laliwala2006semantic}
Laliwala, Z., Sorathia, V., Chaudhary, S.: Semantic and rule based event-driven
  services-oriented agricultural recommendation system. In: 26th IEEE
  International Conference on Distributed Computing Systems Workshops
  (ICDCSW'06). pp. 24--24. IEEE (2006)

\bibitem{lee2013pathrank}
Lee, S., Park, S., Kahng, M., Lee, S.G.: Pathrank: Ranking nodes on a
  heterogeneous graph for flexible hybrid recommender systems. Expert Systems
  with Applications  \textbf{40}(2),  684--697 (2013)

\bibitem{middleton2004ontological}
Middleton, S.E., Shadbolt, N.R., De~Roure, D.C.: Ontological user profiling in
  recommender systems. ACM Transactions on Information Systems (TOIS)
  \textbf{22}(1),  54--88 (2004)

\bibitem{minkov2017graph}
Minkov, E., Kahanov, K., Kuflik, T.: Graph-based recommendation integrating
  rating history and domain knowledge: Application to on-site guidance of
  museum visitors. Journal of the Association for Information Science and
  Technology  \textbf{68}(8),  1911--1924 (2017)

\bibitem{mohamed2016multi}
Mohamed, B., Abdelkader, B., M'hamed, B.F.: A multi-level approach for mobile
  recommendation of services. In: Proceedings of the International Conference
  on Internet of things and Cloud Computing. p.~40. ACM (2016)

\bibitem{nissen2018digital}
Nissen, V.: Digital transformation of the consulting industry—introduction
  and overview. In: Digital Transformation of the Consulting Industry, pp.
  1--58. Springer (2018)

\bibitem{mastersthesis}
Pande, C.: Benchmarking Recommender Algorithms for Business Intelligence
  Consultancy. Master's thesis, FHNW University of Applied Sciences and Arts
  Northwestern Switzerland, CH-4600 Olten, Switzerland (2019)

\bibitem{Richter1998}
Richter, M.M.: {Introduction}. In: Lenz, M., Burkhard, H.D.,
  Bartsch-Sp{\"{o}}rl, B., Wess, S. (eds.) Case-Based Reasoning Technology SE -
  1, Lecture Notes in Computer Science, vol.~1400, pp. 1--15. Springer Berlin
  Heidelberg (1998)

\bibitem{tarus2017knowledge}
Tarus, J.K., Niu, Z., Mustafa, G.: Knowledge-based recommendation: a review of
  ontology-based recommender systems for e-learning. Artificial Intelligence
  Review pp. 1--28 (2017)

\bibitem{voorheesBook}
Voorhees, E.M., Harman, D.K.: {TREC -- Experiment and Evaluation in Information
  Retrieval}. The MIT press, Cambridge, Massachusetts (2006)

\bibitem{werth2016self}
Werth, D., Zimmermann, P., Greff, T.: Self-service consulting: conceiving
  customer-operated digital it consulting services. In: Proceedings of SIGOSRA
  2016 (2016)

\bibitem{White2003}
White, S., Smyth, P.: {Algorithms for Estimating Relative Importance in
  Networks}. In: Proceedings of the Ninth ACM SIGKDD International Conference
  on Knowledge Discovery and Data Mining. pp. 266--275 (2003)

\bibitem{witschel15c}
Witschel, H.F., Galie, E., Riesen, K.: {A Graph-Based Recommender for Enhancing
  the Assortment of Web Shops}. In: Proceedings of Workshop on Data Mining in
  Marketing DMM'2015 (2015)

\bibitem{Witschel2018RandomWO}
Witschel, H., Martin, A.: {Random Walks on Human Knowledge: Incorporating Human
  Knowledge into Data-Driven Recommender}. In: Proceedings of the 10th
  International Conference on Knowledge Management and Information Sharing
  (KMIS) (2018)

\bibitem{wu2015fuzzy}
Wu, D., Zhang, G., Lu, J.: A fuzzy preference tree-based recommender system for
  personalized business-to-business e-services. IEEE Transactions on Fuzzy
  Systems  \textbf{23}(1),  29--43 (2015)

\bibitem{yao2015unified}
Yao, L., Sheng, Q.Z., Ngu, A.H., Yu, J., Segev, A.: Unified collaborative and
  content-based web service recommendation. IEEE Transactions on Services
  Computing  \textbf{8}(3),  453--466 (2015)

\bibitem{zhang2012declarative}
Zhang, M., Ranjan, R., Nepal, S., Menzel, M., Haller, A.: A declarative
  recommender system for cloud infrastructure services selection. In:
  International Conference on Grid Economics and Business Models. pp. 102--113.
  Springer (2012)

\bibitem{zhang2018social}
Zhang, Y., Zuo, W., Shi, Z., Yue, L., Liang, S.: Social bayesian personal
  ranking for missing data in implicit feedback recommendation. In:
  International Conference on Knowledge Science, Engineering and Management.
  pp. 299--310. Springer (2018)

\bibitem{zhang2013random}
Zhang, Z., Zeng, D.D., Abbasi, A., Peng, J., Zheng, X.: A random walk model for
  item recommendation in social tagging systems. ACM Transactions on Management
  Information Systems (TMIS)  \textbf{4}(2), ~8 (2013)

\end{thebibliography}
\end{document}